# Interactions between Multipolar Nuclear Transitions and Gravitational Waves

Yao Cheng and Jian Qi Shen[*]
Department of Engineering Physics, Tsinghua University, Beijing 100084, P. R. China
[*]Zhejiang Institute of Modern Physics, Department of Physics, Yuquan Campus, Zhejiang University, Hangzhou 310027, P.R. China
Email: yao@tsinghua.edu.cn

## Abstract

Interactions between multipolar nuclear transitions and gravitational waves (GWs) are theoretically investigated. Two nonclassical scenarios of the GW detection are suggested. We demonstrate in this report that the long-lived Mössbauer nuclides of multipolar transitions are suitable transducers to detect the impinging GWs. Shape deformation and spin flip of nucleus are derived from the Hamiltonian with gravity interaction. The GWs generate a nuclear quadrupole deformation in analogy with the Stark effect, of which the electric field generates the dipole deformation of electron orbits. Likewise in analogy with the nuclear magnetic resonance that the rotating radio-frequency field flips the nuclear spin, the GWs flip the nuclear spin by the resonant helicity-rotation-gravity coupling. The high energy states of the quadrupole deformation in heavy nuclides can dramatically speed up the multipolar transitions, which may be important for the GW detection particularly in the low frequency band, ranging from 100 nHz to 1 Hz.
PACS numbers: 04.80.Nn, 95.55.Ym, 95.55.-n, 23.20.-g

Detection of gravitational waves (hereinafter GWs) has become one of the most essential researches in the beginning of 21st century, which allows us to observe the astronomical activities in a unique way. Plenty of schemes have been proposed [1-7] to search for the extremely weak GWs from the detectable burst occurrences. Major approaches in the last century are mechanical antennas [1], including the broad-band detector utilizing a highly sensitive laser interferometer such as LIGO [2] and the narrow-band resonator such as Weber bar [2]. The anticipated tiny displacement of the test mass ascribed to GWs encounters a major challenge from the background noise. Inspired the historical work of the Hanover geodesic survey by Gauss in 1818, Kawamura and Chen [3] have demonstrated a class of network interferometry for the displacement- and laser-noise-free GW detection. Recently, GW detectors of the Sagnac-type matter-wave interferometry has been developed [4,5], where the geometric Berry phase of loop shall be probed by the coherent matter beam instead of the laser beam. The atomic interferometry has the advantage in many aspects such as small size, high sensitivity, and low frequency [4], which allow us to construct plenty of observatories for fishing the burst occurrences simultaneously and eventually to detect the quasi-stable GW sources in the low frequency band. However, unlike the atomic interferometry, most of the other concepts mentioned above focus on the coherent signal of astronomic sources in the high-frequency band, ranging from 1 Hz to 10 kHz. The cosmic background GWs are random, which are expected in the microwave region. Correlation between two identical GW detectors is thus required and considered at the very high frequency band of 100 MHz [6], in which GWs rotate the polarization of microwave in the process of cyclic propagation. Sorge *et al.* have considered the GW detection by means of a circulating gyroscope [7]. Their result demonstrates that the out-of-plane precession of gyroscope is accumulative and proportional to the algebraic product of the GW frequency, the GW amplitude and the GW phase factor under the resonant condition, when the frequency of gyroscope rotating frame matches one half of the impinging GW frequency.

The aim of the present study is to derive a basic theorem for the simplest and the most direct interactions between GWs and nucleus based on the Borde's work [8]. Two types of actions are investigated, *i.e.* shape deformation and spin flip. The detection sensitivity by the scheme of spin flip is too low for a practical application with the present technology, unless an effective GW amplifier is developed. On the other hand, the detection sensitivity by the scheme of shape deformation is realizable by the collective enhancement of multipolar gamma transition. It sensitivity for experimental realization is possible to implement upon our theoretical estimation obtained from the semiclassical approach of this work. We treat the emitted photons as a messenger to carry out the action of classical GW field on the nuclear transition of interest. The back-action-evading techniques to measure weak GWs [2] are therefore no longer under our concerns. Rough estimations on the detection sensitivity reveal that the long-lived Mössbauer nuclides are GW transducers of great potential in the low-frequency band, ranging from 100 nHz to 1 Hz.

We firstly introduce the detection scheme of spin-flip. The round-trip rotation of a spin-one particle gives the quantum phase of $\Delta\Phi = 2\pi$. Additional Berry phase arises from the non-Euclidean geometry. Phase shift proportional to the GW amplitude can be accumulated in the particular direction allowing easier detection [6]. For simplicity, we consider a particular case of the static external magnetic field $H_z$ aligned with the GW propagation along the z-axis. Nucleus rotates with Larmor frequency

$$\Omega = -\gamma H_z, \tag{1}$$

where $\gamma$ is the nuclear gyromagnetic ratio [9]. In the Larmor rotating frame of $(x', y', z)$, in which the spin $\boldsymbol{\sigma}'$ stands at rest, the rotating observer is free from any force. In other words, the Coriolis force exactly cancels the Lorenz force. Otherwise the internal dynamic of nucleons would give a nonzero net force. In the absence of GWs, spin rotation in the laboratory frame $(x, y, z)$ does not yield any precession out of the rotating plane with

$$\begin{aligned}
\sigma_x &= \sigma_{x'} \cos(\Omega t) + \sigma_{y'} \sin(\Omega t) \\
\sigma_y &= -\sigma_{x'} \sin(\Omega t) + \sigma_{y'} \cos(\Omega t) \\
\sigma_z &= \sigma_z
\end{aligned} \qquad (2)$$

As a matter of fact, nuclear charge center coincide with its mass center. The lattice suspension of nucleus gives neither torque nor friction except for the spin-lattice and the spin-spin relaxations, which can be suppressed by lowering the temperature and reducing the concentration of spin, respectively. The intrinsic spin is then Fermi-Walker transported [1] without flipping. Borde gives the Hamiltonian of particle and GWs in interaction, which is derived from the non-relativistic approach of Dirac equation in the curved space-time [4,8]. We add an extra interaction term with the electro-magnetic (EM) field to the two particular terms of Borde's interaction Hamiltonian, *i.e.* the de Sitter precession, the second term, and the Linet-Tourrenc GW effect, the third term,

$$H_I = \frac{e}{m_{eff} c} \mathbf{A} \cdot \mathbf{P} + \frac{1}{2M} \boldsymbol{\sigma} \cdot \nabla \times \vec{h} \mathbf{P} + \frac{1}{2M} \mathbf{P} \vec{h} \mathbf{P}, \qquad (3)$$

where the first term is EM radiation field of the vector potential $\mathbf{A}$ and the effective mass $m_{eff}$ of proton (hole) in interaction. The other two Borde's terms are for the GW interaction with the total mass $M$ of all nucleons. The GW tensor fields $\vec{h}$ are weak fluctuations deviating from the background Minkowski space metric $\eta_{mn}$, which have two circular polarizations in the TT gauge observed on the rotational $x' - y'$ frame [10],

$$\vec{h} \approx h_{\pm} e^{i\omega_{GW} z/c - i(\omega_{GW} \mp 2\Omega)t} \begin{pmatrix} 1 & \pm i & 0 \\ \pm i & -1 & 0 \\ 0 & 0 & 0 \end{pmatrix} + c.c., \qquad (4)$$

with the light speed in vacuum $c$, the GW frequency $\omega_{GW}$, and the amplitude $h_+$ and $h_-$ for the positive and the negative helicity, respectively. When $\omega_{GW} = 2\Omega$, the Rabi flopping occurs in analogy with the nuclear magnetic resonance excitation by means of the rotating radio-frequency field [9]. In the long wavelength limit of GW, the off-diagonal elements of the spin-rotation-gravity coupling term in eq. (3) are

$$\frac{1}{2M} \langle m' | \boldsymbol{\sigma} \cdot \nabla \times \vec{h} \mathbf{P} | m \rangle = \mp \frac{\omega_{GW} \Omega}{2c} \left( h_{\pm} e^{-i(\omega_{GW} \mp 2\Omega)t} \langle m' | \sigma_{\pm} (x \pm iy) | m \rangle + c.c. \right), \qquad (5)$$

where $\sigma_{\pm}$ are the raising and the lowering operators, respectively. The operator algebra of $P_x = iM\omega_{mm'}x$ has been used. The spin flip is only allowed between states of magnetic quantum numbers $m$ and $m'$ with $|m-m'|=2$ being the selection rule in analogy with the quantum electrodynamics (QED) selection rule of $|m-m'|=1$ for the operator $x \pm iy$ with the z-axis impinging GW. The net spin-flip depends on the occupation numbers between $|m\rangle$ and $|m'\rangle$, i.e. the temperature of nuclear spin at thermal equilibrium. Corresponding to negative and positive spin temperatures [9], the rotation of non-spherical nuclides in resonant interaction with GW collectively generates the in- or out-of-phase GW or EM emissions that flip the nuclear spin by 2 or 1, respectively. The spin-rotation-gravity coupling is impossible for spherical nuclides without the route of spin flip by 2, whereas the EM wave is still allowed. Therefore, according to this selection rule it is possible to verify the graviton spin quanta experimentally. However, the probability is too low due to the small magnitude of $\vec{h}$ tensor in nature. In order to observe a clear signal of the spin-rotation-gravity coupling, it is necessary to amplify the GWs. The non-spherical nuclei prepared at the negative 0⁻ spin temperature [9] may serve as a GW amplifier, if the EM emission is canceled by a suitable arrangement.

Secondly, we emphasize on the enhanced multipolar transitions induced by GWs, which involves three-body interaction among nuclear phonons $\omega_p \sim |\omega_s - \omega_n|$, graviton $\omega_{GW}$ and photon $\omega_\gamma$, under the conditions of $\omega_\gamma \ll \omega_p$, $\omega_{GW} \ll \omega_p$ and $\Omega \ll \omega_p$. The first order perturbation of nuclear wavefunction $|n_1\rangle$ includes the quadrupole transition to intermediate state $|s\rangle$ in the long wavelength limit of GW as,

$$|n_1\rangle = |n\rangle - h_\pm \sum_s \frac{e^{-i(\omega_{GW} \mp 2\Omega)t}|s\rangle\langle s|(P_x \pm iP_y)^2|n\rangle}{2M\hbar(\omega_s - \omega_n - \omega_{GW} \pm 2\Omega)} - h_\pm^\dagger \sum_s \frac{e^{i(\omega_{GW} \mp 2\Omega)t}|s\rangle\langle s|(P_x \mp iP_y)^2|n\rangle}{2M\hbar(\omega_s - \omega_n + \omega_{GW} \mp 2\Omega)} \quad (6)$$

where $h_\pm^\dagger$ are the complex conjugates of GW amplitudes $h_\pm$. This quadrupole deformation is in analogy with the dipole deformation of the well-known Stark effect. However, the major issues discussed here are the multipolarity, the nuclear size and the energy of intermediate states. The nuclear size is several femtometers, about one part in $10^4$ of the atomic size. The quadrupole deformation intrinsically exists, e.g. via collective vibration modes of nuclear phonon, of which energies range from hundreds of keV to MeV [11]. We shall call this phonon-like transition driven by GW as the nuclear Raman effect. The Linet-Tourrenc term [8] $\mathbf{P}\vec{h}\mathbf{P}$ of GW interaction in eq. (3) is quadratic in $\mathbf{P}$, which leads to a significant transition via the intermediate states at the high-energy level. Though the small nuclear size is unfavorable of the long-wavelength GW detection due to the curl operation appearing in the spin-rotation-gravity coupling term, $\boldsymbol{\sigma} \cdot \nabla \times \vec{h}\mathbf{P}/2M$, it is no longer a disadvantage for the Linet-Tourrenc term. The other terms related to the EM interaction such as $\mathbf{A}\vec{h}\mathbf{P}$ and $\mathbf{A}\vec{h}\mathbf{A}$ is not shown in eq. (3) because these are generally smaller than the $\mathbf{P}\vec{h}\mathbf{P}$ term in the weak field limit of $\mathbf{A}$.

Now let us consider the gamma transition in interaction with GWs by applying the second-order time-dependent perturbation theory. Under the above mentioned condition that nuclear phonon energy $\omega_p$ is much larger than the gamma energy $\omega_\gamma$ with $\omega_\gamma \ll \omega_p$, the off-diagonal transition element is expressed in terms of the phonon energy $\omega_p$, corresponding to the energy from the most significant deformation state |s> to |k> $\omega_p \sim |\omega_s - \omega_k|$ or |s> to |n> $\omega_p \sim |\omega_s - \omega_n|$ in the rotational frame as,

$$\langle k_1|\mathbf{A}\cdot\mathbf{P}|n_1\rangle \approx \langle k|\mathbf{A}\cdot\mathbf{P}|n\rangle$$
$$+h_\pm^\dagger e^{i(\omega_{GW}\mp 2\Omega)t}\frac{M}{2\hbar}\omega_p\left[\langle k|\mathbf{A}\cdot\mathbf{P}|s\rangle\langle s|(x\mp iy)^2|n\rangle+\langle k|(x\mp iy)^2|s\rangle\langle s|\mathbf{A}\cdot\mathbf{P}|n\rangle\right] \quad (7)$$
$$+h_\pm e^{-i(\omega_{GW}\mp 2\Omega)t}\frac{M}{2\hbar}\omega_p\left[\langle k|\mathbf{A}\cdot\mathbf{P}|s\rangle\langle s|(x\pm iy)^2|n\rangle+\langle k|(x\pm iy)^2|s\rangle\langle s|\mathbf{A}\cdot\mathbf{P}|n\rangle\right].$$

This treatment is standard in QED known as the second-order radiative transition [12]. The additional two terms in eq. (7) give rise to the speed-up transition corresponding to the original gamma transition of $\langle k|\mathbf{A}\cdot\mathbf{P}|n\rangle$ without the presence of GW. Here, we select $|\Delta J|=2$ for the most significant term of the GW-induced quadrupole deformation of same parity. The originally forbidden multipolar gamma transition from $|k\rangle$ to $|n\rangle$ speeds up via the intermediate state $|s\rangle$. For example, the E3 octupole transition of the particular $\Delta J=3$ case speeds up via the E1 dipole transition with $\Delta J=1$ from the state $|s\rangle$ by a factor of $10^9 A^2 E_p^2 E_\gamma^{-4} h_\pm^2$ in units of MeV (see appendix), which is derived from the Weisskopf estimates [11], with the atomic mass number $A$, with the energy $E_p=\hbar\omega_p$ of the virtual phonon transition, and the gamma energy $E_\gamma=\hbar\omega_\gamma$. This formula demonstrates that the low photon energy, the heavy nuclides and the high-energy quadrupole deformation are favorable for the GW detection.

The transition enhancements for [235m]U [13], [103m]Rh, [93m]Nb and [45m]Sc according to eq. (7) have been estimated in table 1. These transitions drastically speed up, however, they are still far too small to be regularly observed, even for one of the brightest GW sources known as RX J0806.3+1527, a pair of white dwarfs orbiting with period of 321 seconds and strain amplitudes of $h_\pm \sim 10^{-21}$ [14]. Nevertheless, the Mössbauer nuclei in a thick crystal are capable of coherently speeding up the decay further. In particular, the superradiance is enhanced by a factor N by the multibeam Borrmann channel of enhanced transmission [15], in which N is the number of nuclei in the crystal coherently trigged by the long-wavelength GWs. To give the observable superradiance of [93m]Nb for the GW of $h_\pm \sim 10^{-21}$, it requires N~$10^{21}$ to suppress the incoherent internal conversion [15]. The oscillation on K x-rays emitted from the atomic shell of [93m]Nb become observable in accordance with the GW frequency $\omega_{GW}$.

Three Mössbauer nuclides of concern, *i.e.* $^{103}$Rh, $^{93}$Nb, and $^{45}$Sc, are 100% in natural abundance. The lattice of resonant nuclei becomes a photonic crystal [16], if the gamma dissipation by shell electrons is sufficiently weak, *i.e.* transparent, in the multibeam Borrmann channel. The anomalous emission of an internal Mössbauer emitter of multipolarity near the band edge has long been predicted by Hannon and Trammel [15]. By matching the gamma wavelength with the lattice constant, the photon propagating via Bragg reflected paths constructively interferes leading to the anomalous emission, which is indeed related the localization of superradiance described by John and Quang [16]. The total number of the nuclei, N, participating in the formation of the photonic crystal refers to those of whole crystal by internal emission [16]. This is very different from the number of crystalline layers corresponding to the crystal thickness for the nuclear forward scattering created by the external source [15]. The wavelength of the Mössbauer photon is of the nuclear nature, while the lattice constant is of the atomic nature. In reality, it is difficult to match these two natures, unless a collective mode of the multi-photon transition in resonance with the lattice exists.

To conclude, we suggest two novel schemes for GW detection. The major scheme of interest is based on the interactions between multipolar nuclear transitions and GWs, particularly in the low-frequency band as the supplement to the high-frequency GW detector such as LIGO. One of the remarkable features of the present scheme is that the nuclear shape deformation caused by the impinging GWs speeds up the multipolar gamma transitions of heavy nuclides. As the GWs are capable of flipping the nuclear spin, the graviton spin quantization can then be verified experimentally by suppressing the spin flip due to the EM force. The effects presented here is potentially important in experimental general relativity.

We acknowledge Yanhua Shih at Maryland University for the useful discussion. This work is supported by the NSFC grant 10675068.

| Isotope | $E_\gamma$ | Multipolarity | $\lambda_W \left( s^{-1} \right)$ | α | $\lambda \left( s^{-1} \right)$ | $E_s$ | $\Delta\lambda/\lambda_w$ |
|---|---|---|---|---|---|---|---|
| $^{103m}$Rh | 39.8 keV | E3 | $6\times10^{-5}$ | ~1350 | $2\times10^{-4}$ | 0.9 MeV | $3\times10^{18} h_\pm^2$ |
| $^{93m}$Nb | 30.8 keV | M4 | $8\times10^{-16}$ | ~$2\times10^5$ | $1\times10^{-9}$ | 14 MeV | $2\times10^{21} h_\pm^2$ |
| $^{235m}$U | 76.8 eV | E3 | $3\times10^{-23}$ | ~$10^{21}$ | $5\times10^{-4}$ | 1 MeV | $2\times10^{30} h_\pm^2$ |
| $^{45m}$Sc | 12.4 keV | M2 | 0.1 | ~350 | 0.3 | 8 MeV | $10^{14} h_\pm^2$ |

Table 1: Enhanced decay via GW-induced quadrupole deformation. $E_\gamma$: Gamma energy; $\lambda_W$: Weisskopf estimates of transition rate in s$^{-1}$ [11]; α: Estimated total internal conversion coefficient [11]; $\lambda$: Measured transition rate in s$^{-1}$ [11]; $E_s$: Highest reported intermediate state of quadrupole deformation [11]; $\Delta\lambda/\lambda$: Transition enhancement due to GWs, which are calculated from the Weisskopf estimates in appendix. The three nuclides of $^{103}$Rh, $^{93}$Nb, $^{45}$Sc are isotopes of 100% natural abundance.

## Appendix:

According to eq. (7), the speed-up transition probability Δλ in unit of s$^{-1}$ form |n$_1$> to |k$_1$> is estimated by the Fermi golden rule known as the Weisskopf estimates [11] evaluated as,

$$\Delta\lambda(k_1 \to n_1) \sim \frac{2(L'+1)}{L'[(2L'+1)!!]^2}\left(\frac{3}{L'+3}\right)^2 \frac{e^2}{\hbar}\kappa^{2L'+1}R^{2L'}\left(\frac{M}{2\hbar}\omega_p R^2 h_\pm\right)^2, \quad (A-1)$$

where $\kappa = \omega_\gamma/c$ is the wavevector of gamma, R=A$^{1/3}$R$_0$ is the nuclear radius with R$_0$=1.26fm, the atomic mass number A, and the frequency $\omega_p$ for the transition from the most significant intermediate state |s> to the ground state |n> or to the Mössbauer state |k>. According to eq. (7), the reduced angular momentum L' is |n-s| or |k-s|. The original transition probability λ according to the Weisskopf estimate in units of s$^{-1}$ form |k> to |n> is

$$\lambda(k \to n) = \frac{2(L+1)}{L[(2L+1)!!]^2}\left(\frac{3}{L+3}\right)^2 \frac{e^2}{\hbar}\kappa^{2L+1}R^{2L}, \quad (A-2)$$

where the angular momentum is L=|n-k|. The multipolar transition of E3 speeds up from the E1 transition driven by GW's in units of 1 MeV written as,

$$\frac{\Delta\lambda_{E1}(E_\lambda)}{\lambda_{E3}(E_\lambda)}\cdot\left(\frac{M}{2\hbar}\omega_p R^2 h_\pm\right)^2 \sim 10^9 A^2 E_p^2 E_\gamma^{-4} h_\pm^2. \quad (A-3)$$

The multipolar transition of M4 speeds up from the M2 transition in units of 1 MeV written as,

$$\frac{\Delta\lambda_{M2}(E_\lambda)}{\lambda_{M4}(E_\lambda)}\cdot\left(\frac{M}{2\hbar}\omega_p R^2 h_\pm\right)^2 \sim 2.8\times 10^9 A^2 E_p^2 E_\gamma^{-4} h_\pm^2. \quad (A-4)$$

The multipolar transition of M2 speeds up from the E1 transition in units of 1 MeV written as,

$$\frac{\Delta\lambda_{E1}(E_\lambda)}{\lambda_{M2}(E_\lambda)}\cdot\left(\frac{M}{2\hbar}\omega_p R^2 h_\pm\right)^2 \sim\sim 10^3 A^{10/3} E_p^2 E_\gamma^{-2} h_\pm^2. \quad (A-5)$$